# Optical variability of the supersoft source RX J0019.8+2156

J. Greiner[1], W. Wenzel[2]

[1] Max-Planck-Institut für Extraterrestrische Physik, 85740 Garching, Germany, jcg@mpe-garching.mpg.de
[2] Sternwarte Sonneberg, 96515 Sonneberg, Germany

**Abstract.** We present a 100 yr optical lightcurve of the recently discovered supersoft X-ray source RX J0019.8+2156 as deduced from photographic plates of Harvard and Sonneberg Observatory. Apart from the periodic orbital variations two different timescales of optical variability are discovered. The timescales and amplitudes of this variability are discussed in the framework of the steady nuclear burning model.

**Key words:** supersoft X-ray sources – white dwarfs – binaries – variable stars

## 1. Introduction

After the discovery observations of supersoft X-ray sources (SSS) with *Einstein*, the *ROSAT* satellite has discovered two dozens new SSS (Greiner et al. 1991, Schaeidt et al. 1993, Greiner et al. 1994, Motch et al. 1994, Kahabka et al. 1994, Beuermann et al. 1995, Supper et al. 1995). The *ROSAT* spectra of SSS are well described by blackbody emission of very low temperature (kT≈ 25–40 eV) and a luminosity close to the Eddington limit (Greiner et al. 1991). Several SSS exhibit X-ray variability. The first of the new *ROSAT* discovered SSS – RX J0527.8–6954 – must have been a factor of ten fainter during the earlier *Einstein* observations when it was in the FOV but not detected. A transient SSS was observed during the *ROSAT* All-Sky-Survey to rise from below 0.06 cts/sec to about 2 cts/sec within a few days (Schaeidt et al. 1993). Later observations showed the source being alternately on and off on a few months timescale (Schaeidt et al. 1995). These observations suggest that the variability occurs on timescales which are short compared to the nuclear burning timescale of novae or the limit cycle length as proposed by van den Heuvel et al. (1992).

RX J0019.8+2156 was found during the search for galactic SSS in the *ROSAT* All-Sky-Survey data and optically identified with a V=12.2 mag object (Beuermann

*Send offprint requests to*: J. Greiner

et al. 1995). Its optical spectrum is very similar to that of CAL 83 (Cowley et al. 1984). As the optically brightest SSS, RX J0019.8+2156 is best suited for a study of the long-term optical behaviour of this class of objects.

## 2. Observational results

### 2.1. Photographic plates

The Harvard plate collection comprises plates taken at different observatories and with different instrumentations (catalogued as a specific series each). Plates of the series I, RH, A, AC, MC, and Damon were used for this study (see Table 1). From the Sonneberg plate collection the blue sensitive plates of the sky patrol (SHÜ) have been used. They were gained in general with Zeiss Ernostar and Tessar type cameras of focal lengths between approximately 20 cm and 30 cm. The majority of plates used here are centered at $0^h+20°$. Our comparison stars (Fig. 1) have been linked to the system of the Mt. Wilson International magnitudes of Selected Area No. 68 (Seares et al. 1925).

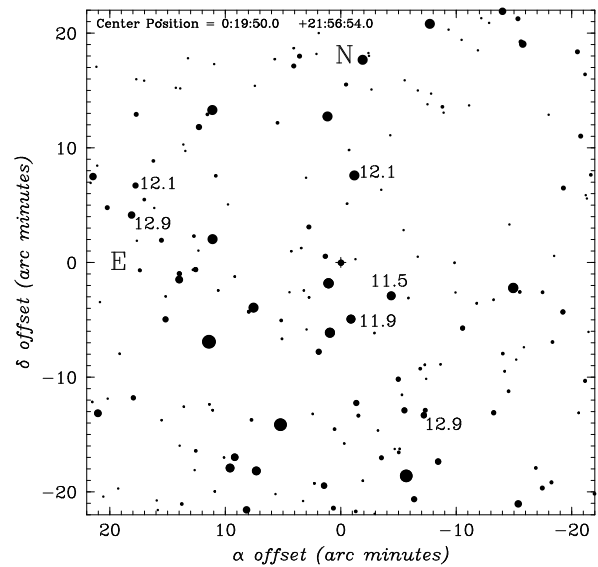

**Fig. 1.** Finding chart of RX J0019.8+2156 (cross in the center) with the photographic magnitudes of the comparison stars.

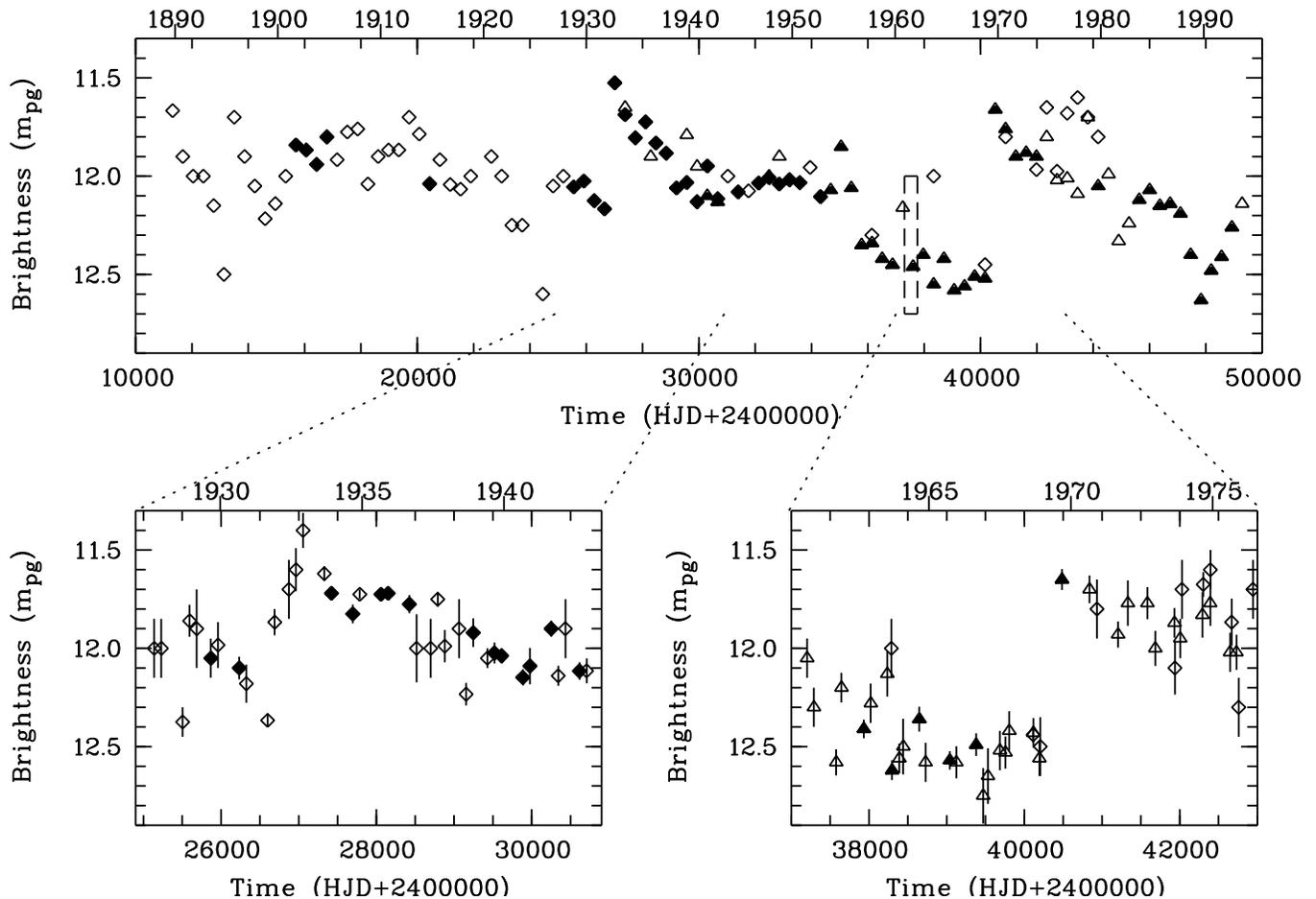

**Fig. 2.** Optical lightcurve of RX J0019.8+2156 derived from photographic plates at Harvard (lozenges) and Sonneberg (triangles) Observatory, respectively. Each data point represents the mean of one season. Filled symbols contain more than 9 individual measurements with 1$\sigma$ statistical errors being less than 0.05 mag, whereas open symbols contain 9 measurements or less. The dashed box marks the part shown in more detail in Fig. 3. The lower panels are blow-ups of the two time intervals covering the sudden intensity jumps. Here, the data are shown in bins of three months together with 1$\sigma$ error bars.

The use of different emulsions in the past 100 years together with the existence of strong emission lines in the optical spectrum of RX J0019.8+2156 demanded some care in the data reduction. Thus, plates with unknown emulsion type or filter combination have been discarded. In addition, several other stars in the vicinity of RX J0019.8+2156 have been monitored to ensure the recognition of rare, falsely documented information on emulsion and/or filter. We estimate the error of a single magnitude determination to be 0.15 mag.

### 2.2. Optical variability

In addition to the optical variability already described in Beuermann et al. (1995), namely

1. a cyclic variation of a period of 15.8 hours with an amplitude of about 0.3 mag,
2. quasiperiodic "pulsations" of roughly 2 hours length each and a range of <0.1 mag,

we detected two further components of variability:

3. long-term variations, seemingly non-periodic, with timescales of up to 20 years and an amplitude of 1 mag (Fig. 2) and
4. irregular fluctuations of timescales of weeks to months and small range (Fig. 3).

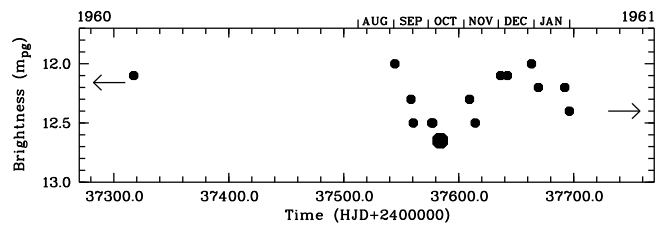

**Fig. 3.** Typical variation on the timescale of weeks. The large symbol represents 9 magnitude determinations, the remaining are single measurements (all Sonneberg plates). Data of the phase interval 0.9–0.1 have been discarded. Arrows mark the mean of the preceding and following seasons.

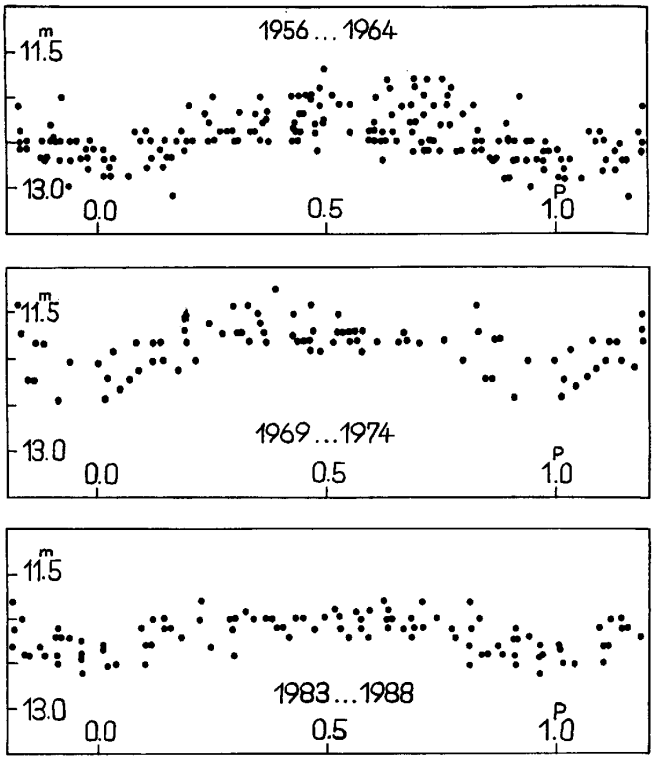

Fig. 4. Orbital variation of RX J0019.8+2156 in three time intervals. The period (as given in the text) is stable at least over the last 40 years.

Table 1. Archival plates used

| Plate series | $m_{lim}$ (mag) | Time span | No. of plates |
| --- | --- | --- | --- |
| Harvard I | 17 | 1889–1942 | 53 |
| Harvard A | 17 | 1898–1936 | 4 |
| Harvard AC | 13 | 1898–1952 | 379 |
| Harvard MC | 17 | 1914–1979 | 18 |
| Harvard RH | 15 | 1928–1963 | 173 |
| Harvard DB | 15 | 1968–1979 | 40 |
| Sonneberg SHÜ | 14 | 1931–1994 | 689 |

There are two large optical brightness jumps by about 1 magnitude. The rise time of the jump in 1969 is shorter than 10 months, while the jump in 1931/32 has a rise time of 14 months. Both of these intensity jumps are followed by a slow relaxation continuing over ≈20 years. Whether or not there was a similar jump at the end of the last century (which would imply a nearly periodic recurrence time scale of ≈40 years) cannot be decided due to the sparse coverage. But certainly, at these times the object also showed the full amplitude of variability (>1 mag).

Many of the seasonal means as plotted in Fig. 2 are averages over irregular fluctuations as shown in Fig. 3. These variations occur on timescales of weeks to months, and the variability pattern are different from year to year. Thus, a considerable part of the scattering in Fig. 2 can be assigned to these irregular, short-term fluctuations with the resulting mean being dependent on the different coverage of bright and faint episodes.

In the best-covered sections of the Sonneberg material, when also emulsion types and cameras used were homogeneous, component 1 can be easily traced despite of the other components. Our observations present no contradictions to the period given by Beuermann et al. (1995). We were, however, able to substantially improve it and derived the following lightcurve elements (ephemeris):

Min.(hel.) = 243 5799.247 + 0.$^d$6604565 × E

which is valid at least for the years 1955–1993 (Fig. 4).

## 3. Discussion

### 3.1. Binary dimension

The stability of the period over many years is most easily explained by orbital motion. The very broad minimum, and the obvious vanishing of the "pulsational" feature (above item 2) during the central parts of the minimum point towards an eclipse of a bright accretion disk. The presence of a strong reflection effect which would show the same phase behaviour as an eclipse phenomenon, cannot be excluded by our observations alone, but is not probable in view of the spectral behaviour (Beuermann et al. 1995). The dimension of the circular orbit – half axis a = $P^{2/3}(M_1+M_2)^{1/3}$ – fits well with the half amplitude of the observed radial velocity curve – v = $2\pi((M_1+M_2)/P)^{1/3}$ giving a ≈ 4 solar radii, $M_1+M_2$ ≈ 2 solar masses. We conclude therefore that one of the most promising models for the object is an accreting white dwarf (WD) with nuclear burning in a classic cataclysmic system. We note that models comprising a high mass component or a giant secondary star (classic symbiotic binary) can be excluded.

### 3.2. The long-term lightcurve

To variable star experts long-term lightcurves with timescales of several years and amplitudes of the order of 1 mag are not uncommon. Shell stars of the Pleione type (see Hoffmeister et al. 1985 for composite lightcurves) or slowly varying red giants and supergiants (ibid) exhibit such behaviour and the objects of the present kind could be mistaken for them if only photometric observations were at hand. However, we also know various long-term phenomena in mass transferring binaries. Examples are long standstills between eruption phases (Z Cam) or in outburst (UX UMa), inactive states in LMXBs (HZ Her), long lasting bright or faint episodes in "nova-like" variables (KR Aur, MV Lyr, TT Ari), or a slowly varying minimum brightness in dwarf novae (V 426 Oph). In most of these cases a long-term change of the mass transfer emerging from the secondary is the simplest explanation.

RX J0019.8+2156 had an X-ray intensity of of 2.1 cts/sec during the All-Sky-Survey (1990), and 2.0 cts/sec in pointed *ROSAT* observations in 1992 and 1993 (Beuermann et al. 1995), i.e. there was no variation in X-ray luminosity over 3 years. Also, no temperature changes could be found. During the same time interval the mean optical intensity has increased by nearly 0.5 mag. Unfortunately, there are no exactly simultaneous *ROSAT* and optical measurements. Given the short-term variability on timescale of weeks (Fig. 3), we cannot draw a firm conclusion on the relation of X-ray and optical luminosity.

*3.3. Burning white dwarfs*

The most popular model involves steady nuclear burning on the surface of an accreting WD (van den Heuvel et al. 1992). Hydrogen burning on the WD surface is stable for $\dot{M} \geq 10^{-7}\ M_\odot/\mathrm{yr}$ (Paczynski & Zytkow (1978). However, the burning timescale (100–300 yrs) is much longer than the observed optical variability of RX J0019.8+2156. Possibly, the observations can be understood by temporarily reduced mass transfer which leads to long-term "oscillations" between stable and unstable H burning. Consequently, the optical brightness jumps by 1 magnitude could be caused be ignitions of H (or He) burning: Extensive steady state calculations of massive white dwarfs accreting at high rates have shown (Iben 1982, Fujimoto 1982) that for WD masses in excess of 1.0 $M_\odot$ the on/off timescales of H flashes become shorter than 50 years for accretion rates of the order of $10^{-7}\ M_\odot/\mathrm{yr}$. Hydrodynamic calculations have shown (Livio et al. 1989) that the luminosity in the off state increases with increasing $\dot{M}$, thus resulting in smaller outburst amplitudes. Thus, one could imagine that (possibly varying) mass transfer near the burning rate drives the WD only into mild flashes. After having burned the rather small hydrogen layer, the system returns to instability. In order to avoid large amplitude flashes, the subsequent cooling (along the constant radius track in the HR diagram) has to be stopped by resuming accretion at a high rate. Thus, hydrogen ignition on massive WDs would not necessarily be connected with large amplitude intensity variations.

An interesting complication in the above scenario is that though no mass is ejected at accretion rates of about $10^{-7}\ M_\odot/\mathrm{yr}$ the WD expands to $\approx 80\ R_\odot$ (Livio et al. 1989) being much larger than the binary dimension. Sion & Starrfield (1994) therefore proposed to consider extremely hot but low-mass WDs ($\approx 0.5$–$0.7\ M_\odot$) which remain at compact radii ($R_{WD} \leq 0.1\ R_\odot$ for $10^{-8}\ M_\odot/\mathrm{yr}$). However, it remains unclear how the intensity variations and amplitudes as observed in RX J0019.8+2156 can be explained. Certainly, shell flashes in such systems are much too violent, and the deduced cycle lengths are a few thousand years (ibid).

The above scenario has attributed the cause of the variability to changing mass transfer rates which in turn lead to variable nuclear burning rates and hence departures from a limit cycle recurrence time. Consequently, one should also ask for the possible reasons of mass transfer variations. First, if the donor star is more massive than the compact object, mass transfer is unstable on a thermal timescale. Second, if indeed a F or G type dwarf star (as suggested for CAL 83 and CAL 87) is the mass sponsoring secondary component, our hypothesis is that variable median and long-term intrinsic activity of the secondary ("solar cycle activity", depending on the size of the convective atmosphere) might give rise to a changing mass transfer, and therefore to an optical variability. Spectroscopic investigation of the mass transfer and accretion rate during low and high states can clarify this point. Third, also intensity and/or spectral variations of the irradiation of the secondary can affect the mass transfer.

## 4. Conclusion

In addition to the optical variability of RX J0019.8+2156 reported in Beuermann et al. (1995) we have discovered two further timescales of variability. We believe these to be consequences of unstable mass transfer onto a WD which sporadically is burning hydrogen. However, some of the details of the proposed models are not in line with the observed amplitude and timescale of variability. In addition, the applicability of some of the previously applied assumptions such as accretion via a disk instead of spherical accretion, the effect of the expected strong radiation driven wind, the angular momentum loss carried by the escaping gas and tidal effects in these tight binaries might be explored more carefully.

*Acknowledgements.* We are extremely grateful to K. Beuermann for communicating the SSS nature and identification of RX J0019.8+2156 prior to publication, which only enabled the timely work presented here. JG is indebted to M. Hazen for support at the Harvard plate stack, and to I. Iben for an enlightening discussion on unstable H burning. We thank C. Motch for the plot software for Fig. 1 and the anonymous referee for helpful comments. JG and WW are supported by the Deutsche Agentur für Raumfahrtangelegenheiten (DARA) GmbH under contract Nos. FKZ 50 OR 9201 and 05 5S 0414.

## References

Beuermann K., et al. 1995, (accompanying letter)
Cowley A.P., Crampton D. et al. 1984, ApJ 286, 196
Fujimoto M.Y., 1982, ApJ 257, 767
Greiner J., Hasinger G., Kahabka P. 1991, A&A 246, L17
Greiner J., Hasinger G., Thomas H.-C., 1994, A&A 281, L61
Hoffmeister C., Richter G., Wenzel W., 1985, Variable Stars, Springer, Heidelberg
Iben I., 1982, ApJ 259, 244
Kahabka P., Pietsch W., Hasinger G., 1994, A&A 288, 538
Livio M., Prialnik D., Reger O., 1989, ApJ 341, 299
Motch C., Hasinger G., Pietsch W., A&A 284, 827
Paczynski B., Zytkow A., 1978, ApJ 222, 604
Schaeidt S., Hasinger G., Trümper J., 1993, A&A 270, L9
Schaeidt S., et al. 1995, (in prep.)
Seares F., Joyner M.C., Richmond M.L., 1925, ApJ 61, 303
Sion E.M., Starrfield S.G., 1994, ApJ 421, 261
Supper R., et al. 1995 (in prep.)
van den Heuvel E., Bhattacharya D. et al. 1992, A&A 262, 97